\documentclass[oneside, reqno, 12pt]{amsart}

\usepackage{amsmath,amssymb}
\usepackage[dvips]{geometry}

\numberwithin{equation}{section}
\theoremstyle{plain}                
\newtheorem{theorem}{Theorem}[section]
\newtheorem{lemma}[theorem]{Lemma}
\newtheorem{proposition}[theorem]{Proposition}

\theoremstyle{definition}           
\newtheorem{definition}[theorem]{Definition}

\theoremstyle{remark}
\newtheorem{remark}[theorem]{Remark}

\newcommand{\argmax}{\operatorname{argmax}}

\providecommand{\R}{} \renewcommand{\R}{{\mathbb R}}

\def\R{\mathbb{R}}

\def\R{\mathbb{R}}

\makeatletter 
\@addtoreset{equation}{section}

\makeatother 
\begin{document}
\begin{center}
{\Huge\bf Investment and Consumption
 without Commitment}
\mbox{}\\
\vspace{2cm}
Ivar Ekeland
\footnote{Work
supported by PIMS under NSERC grant 298427-04.}\\
Department of Mathematics\\
The University of British Columbia\\
Vancouver, BC, V6T1Z2\\
ekeland@math.ubc.ca\\

\vspace{1cm}
Traian A. Pirvu\\
Department of Mathematics\\
The University of British Columbia\\
Vancouver, BC, V6T1Z2\\
tpirvu@math.ubc.ca\\
\vspace{1cm}

\mbox{}\\

\today
\end{center}

\noindent {\bf Abstract.}
In this paper, we investigate the Merton portfolio management problem in the context
of non-exponential discounting. This gives rise to time-inconsistency of the decision-maker. If the decision-maker at time $t=0$ can commit his/her successors, he/she can choose the policy that is optimal from his/her point of view, and constrain the others to abide by it, although they do not see it as optimal for them. If there is no commitment mechanism, one must seek a subgame-perfect equilibrium strategy between the successive decision-makers. In the line of the earlier work by Ekeland and Lazrak \cite{EkeLaz} we give a precise definition of equilibrium strategies in the context of the portfolio management problem, with finite horizon, we characterize it by a system of partial differential equations, and we show existence in the case when the utility is CRRA and the terminal time T is small. We also investigate the infinite-horizon case and we give two different explicit solutions in the case when the utility is CRRA (in contrast with the case of exponential discount, where there is only one). Some of our results are proved under the assumption that the discount function $h(t)$ is a linear combination of two exponentials, or is the product of an exponential by a linear function.

\vspace{1cm}


\begin{quote}

\end{quote}

\begin{flushleft}
\end{flushleft}

\setcounter{equation}{0}
\section{Introduction}

The discounted utility model (DU) has been in use since the beginning of economic theory. Landmark papers ar the ones by Ramsey in $1928$ and Samuelson in $1937$. There is by now a very rich literature, the common assumption being that  
the discount rate is constant over time so the discount function is
exponential. The model makes it possible to compare outcomes occurring at different
times by discounting future utility by some constant factor. A decision maker
with high discount rates exhibit more impatience (care more about the future)
than one with low discount rates. Most of financial-economics works have
considered that the rate of time preference is constant (exponential discounting).
 However there is growing evidence to suggest that this may not be the case.
Ainslie \cite{Ain} performed empirical studies on human and animal behavior
and found that discount functions are almost hyperbolic. Loewenstein and Prelec
\cite{LoPre} show four drawbacks of exponential discounting and propose a model which
accounts for them. They discuss implications for savings behavior and estimation of
discount rates.

As soon as discounting is non-exponential, the decision-maker becomes time-inconsistent: a policy, to be implemented after time $t>0$, which is optimal when discounted at time $0$, no longer is optimal if it is discounted at a later time, for instance $t$. If the decision-maker at time $0$ can commit the later ones, that is, constrain them to follow the policy he/she has decided upon, then the policy which is optimal from his/her perspective can be implemented. But, apart for the dubious validity of doing that (why should his/her perspective be better than the one of decision-maker at time $t$, who, after all, will have to carry out a policy decided long before, and will be the one to bear its consequences ?) it is not often the case that management decisions are irreversible; there will usually be many opportunities to reverse a decision which, as times goes by, seems ill-advised.  

{\bf{Existing Research}}

Dynamic inconsistent behavior was first formalized analytically by Strotz \cite{STRO}.
Further work by Pollak \cite{Pol}, Peleg and Yaari \cite{PeYa}, Goldmann \cite{Gol}
on this issue advocates that the  
policies to be followed should be the output of an intra-personal game among different temporal selves
(today's self is a different player from tomorrow's self). Laibson \cite{Lai} considers
a discrete time consumption-investment economy without uncertainty. An agent (self)
 observes past consumption and financial wealth levels and chooses the consumption level
for period $t.$ They establish existence of a unique subgame perfect equilibrium. It is
characterized by time-dependent consumption rules which are linear in wealth and should
satisfy an Euler type equation. The work of Barro \cite{Bar} is in a non-stochastic
Ramsey model with logarithmic utility, and the discount function as a special form, whereby the rate of time preference is high in the near future but almost constant in the distant future.  Krusell and Smith \cite{KruSm} also consider a discrete non-stochastic Ramsey paradigm with quasi-geometric discounting. They seek the equilibrium policy as the solution of a subgame-perfect equilibrium where the players are the agent and her future selves, and they show that there are several solutions to this problem. All this work is in the deterministic case, usually with discrete time. 

Ekeland and Lazrak \cite{EkeLaz} consider a deterministic problem with continuous time, namely the Ramsey problem of economic growth with non-exponential discounting. They define an subgame perfect equilibrium strategies by letting the decision maker at time $t$ build a coalition
with his/her immediate successors $s,$ with $s\in[t,t+\epsilon]$ and by letting $\epsilon\rightarrow0$. They show that these strategies are
characterized by two equivalent equations: a partial differential equation with a non-local
term and an integral equation. The PDE coincide with the Hamilton-Jacobi-Bellmann equation
of optimal control if the discounting is exponential.

{\bf{Our Contributions}}

Dynamic asset allocation has been an area in finance which received a lot of attention
lately. The ground-breaking paper in this literature is Merton \cite{Mer71}.
He considered a model consisting of a risk-free asset with constant rate of return
and one or more stocks, each with constant mean rate of return and volatility.
An agent invests in this market and consume to maximize her expected utility
of intertemporal consumption and final wealth. Merton was able to derive closed form
solution for Constant Relative Risk Aversion (CRRA) and Constant Absolute Risk Aversion
(CARA) preferences. It turns out for CRRA preferences, the optimal consumption and
investment in the risky asset are a constant proportion of wealth. This is not the case
for CARA although they are still linear in wealth. Karatzas et al. \cite{KarLehShr87},
Cox and Huang \cite{CoxHua89} also solved the optimal
investment and consumption problem by the static martingale method. In another paper Merton \cite{Mer69} also investigates the
infinite horizon case. Although it yields the same quantitative results it is easier to handle -
provided the constant discount rate satisfies a condition known in the literature as the transversality condition. It does not seem to have much empirical support, and what happens if it is not satisfied is a matter of debate.

All the papers mentioned above are in the exponential discounting paradigm which as we
shown earlier was challenged by some economic literature.

The goal of this paper is to understand how non-exponential discounting affects
an agent's investment-consumption policies in a Merton model. First we show that
doing naive optimization does not work in the absence of commitment technologies.
Then we follow the approach of \cite{EkeLaz} and introduce the concept of equilibrium policies in
 a stochastic context. In the finite horizon case we give a first description
 of the equilibrium policies for a general discount function through the solutions
of a flow of BSDEs. If the discount function is exponential, these conditions reduce to the classical HJB equation, so the equilibrium policy coincide with the optimal one given by dynamic programming. We then introduce special classes of discount functions, type I which is a linear combination of two exponentials, and type II which is the product of an exponential by a polynomial of degree one, and we show that for discount functions in that class, equilibrium solutions can be characterized by a system of HJB equations.
We go one step further and characterize them through a parabolic PDE system.  
This system does not seem to have been considered before. Existence of a solution can
be established in the case of CRRA utility.

It is perhaps surprising that the same integral
equation (IE) from \cite{EkeLaz} holds in this model for the special discounts. This suggests
that (IE) holds for any discounts. We can see this for other types of discount but we are
not concern with a general result in this direction. In an infinite horizon model the novel feature is
stationarity and the key observation is, to borrow a sentence from \cite{EkeLaz},
the decision maker at time $t$ resets her watch so that time $s$ becomes $s-t,$ so she faces
the same problem as the decision-maker at time $0.$ Keeping this in mind we define the equilibrium
policies in this context. We follow the same approach as for finite horizon: first we describe them
through the solutions of a infinite horizon BSDE and then for the special discounts through an ODE
system and an integral equation. We end this case with the study of CRRA preferences were we find
interesting results. As one might probably expected the equilibrium policies and optimal ones coincide
if the discounting is exponential given the transversality condition of \cite{Mer69}. However we find
a weaker condition to grant existence of an equilibrium policy so there may be circumstances when the
equilibrium policy exists but the optimal one may not (in the sense that one can not perform the verification
argument). Even more interesting for type I discounts we find instances when there are two equilibrium policies.
 Moreover transversality-like conditions were worked out. The equilibrium policies for CRRA preferences resemble
 the optimal ones. More precisely the agent invests the same constant proportion of her wealth in the risky asset as in the
 Merton problem and consumes also a constant proportion of her wealth which is different than the Merton one for non-exponential
 discounting.

{\bf{Organization of the paper}}

The reminder of this paper is organized as follow.
  In section $2$ we describe the model and formulate the objective. Section $3$ and
   $4$ treat the finite horizon and infinite horizon problem. The conclusions are summarized
in section $5.$ The paper ends with
   Appendix containing the proofs.

\section{The model and problem formulation}

\subsection{Financial Market}
 We adopt a model for the financial market consisting of a saving account and one stock (risky asset).
 The inclusion of more risky assets can be achieved by notational changes. The saving account accrues
interest at the riskless rate $r>0.$ The stock price per share follows an exponential Brownian motion
 $$
dS(t)=S(t)\left[\alpha\,dt
+\sigma\,dW(t)\right],\quad0\leq
t\leq\infty,
$$
where $\{W(t)\}_{t\in[0,\infty)}$ is a $1-$dimensional Brownian motion on a filtered probability space
$(\Omega,\{\mathcal{F}_t\}_{0\leq t\leq T},\mathcal{F},P).$ The filtration
$\{\mathcal{F}_t\}_{0\leq t\leq T}$ is the completed filtration generated by $\{W(t)\}_{t\in[0,\infty)}.$
As usual $\alpha$ is \textit{ the mean rate of return}
and $\sigma>0$ is {\em the volatility}. Let us denote
$\mu\triangleq\alpha-r>0$ {\em the excess return}.

\subsection{Investment-consumption policies and wealth processes}
A decision-maker in this market is continuously investing her wealth in the stock/bond
and is consuming. An investment-consumption policy is determined by the proportion of
current wealth her/she invests in the bond/stock and consume (consumption rate). Formally we have:
\begin{definition}
\label{def:portfolio-proportions}
An $\R^{2}$-valued stochastic process
$\{\zeta(t), c(t)\}_{t\in[0,\infty}$ is called an {\em
  admissible policy process} if
it is progressively measurable, $c(t)\geq 0,\,\,\mbox{for all}\,\,t\in[0,\infty)$ and it
satisfies
\begin{equation}%
\label{kj}
\sup_{0\leq t\leq T }\mathbb{E}[|\zeta(t)|^{m}+c^{m}(t)]<\infty\quad \forall m=1,2,\cdots,
\end{equation}
This condition originates from \cite{Pen} and is needed since we are using their result.
\end{definition}
Given a policy process $\{\zeta(t), c(t)\}_{t\in[0,\infty)}$, the proportion of the wealth 
$X^{\zeta,c}(t),$ at time $t$ invested in the stock is $\zeta(t)$ and the consumption rate is $c(t).$ The
equation describing the dynamics of wealth ${X^{\zeta,c}(t)}$ is given by
\begin{eqnarray}\notag
dX^{\zeta,c}(t)&=&X^{\zeta,c}(t)\left((\alpha\zeta(t)-c(t))\,dt+\sigma\zeta(t)\,dW(t)\right)
+(1-\zeta(t))X^{\zeta,c}(t)r\,dt\\\label{equ:wealth-one}&=&X^{\zeta,c}(t)(\left(r+\mu\zeta(t)-c(t))\,dt 
+\sigma\zeta(t)\,dW(t)\right).
\end{eqnarray}
It simply says that the changes in wealth over time are due solely to gains/loses from investing in stock,
from consumption and there is no cashflow coming in or out. This is usually referred to
as the self-financing condition.

Under the regularity condition \eqref{kj} imposed on $\zeta(t)$ and $c(t)$
above, \eqref{equ:wealth-one},  admits a unique strong solution given by
the explicit expression
\begin{equation}\notag%
    \begin{split}
      X^{\zeta,c}(t)=X(0)\exp\left( \int_0^t
        \Big(r+\mu\zeta(u)-c(u)-|\sigma\zeta(u)|^2\Big)\, du+
        \int_0^t \sigma\zeta(u)\, dW(u) \right),
    \end{split}
\end{equation}
The initial wealth $X^{\zeta,c}(0)=X(0)\in (0,\infty)$, is exogenously specified.

\subsection{Utility Function}

All decision-makers have the same von Neumann-Morgenstern utility. This is crucial for understanding the model: time-inconsistency arise, not from a change in preferences, but from the way the future is discounted. All decision-makers try to maximize
the discounted expectation of a function $U:(0,\infty)\rightarrow \mathbb{R}$ strictly increasing and strictly concave,  
which is their (common) utility.  
We restrict  ourselves to utility functions which are continuous differentiable and satisfy the Inada conditions
 \begin{equation}\label{In}
 U'(0+)\triangleq\lim_{x\downarrow 0}U'(x)=\infty,\quad U'(\infty)\triangleq\lim_{x\uparrow{\infty}}U'(x)=0.
 \end{equation}
We shall denote by $I(\cdot)$ the (continuous, strictly decreasing) inverse of the marginal utility function $U'(\cdot),$
and by \eqref{In}
\begin{equation}\label{In1}
 I(0+)\triangleq\lim_{x\downarrow 0}I(x)=\infty,\quad I(\infty)\triangleq\lim_{x\uparrow{\infty}}I(x)=0.
\end{equation}
The agent is deriving utility from intertemporal consumption and final wealth. Let $U$ be the utility of
intertemporal consumption and $\hat{U}$ the utility of the terminal wealth
at some non-random horizon $T$ (which is a primitive of the model and we may alow it to be infinite).

\subsection{Discount Function}
Unlike other works in this area we do not restrict ourselves to the framework of exponential discounting. Following
\cite{EkeLaz}, a discount function $h:[0,\infty]\rightarrow\mathbb{R}$  is assumed to be continuously differentiable, with:
$$h(0)=1,\,\,h(s)\geq 0 ,$$
and
$$\int_{0}^{\infty}h(t)\,dt<\infty.$$  
The definition and characterization of equilibrium strategies (Theorems \ref{Hamil1} and \ref{Hamil}) hold for general discount functions. We will then particularize them to two special cases, which we will call {\it{pseudo-exponential}} discount functions. They are of two types, type I:
 \begin{equation}
h_{1}(t)=\lambda\exp(-\rho_{1}t)+(1-\lambda)\exp(-\rho_{2}t)
\end{equation}
and type II:
 \begin{equation}
 h_{2}(t)=(1+\lambda t)\exp(-\rho t)
  \end{equation}
  
 Pseudo-exponential discount rates were first considered in the context of time-inconsistency in \cite{EkeLaz}. Note that, in contrast to the more studied case of hyperbolic discount, decision-makers discount the distant future less heavily than the immediate future.  It will be clear from our approach how to handle other cases.

\subsection{Objective}
We conclude this section by formulating our problem. The objective is to find
time consistent policies and the optimal ones may fail to have this feature.
Indeed, if the agent starts with a given positive wealth $x,$ at some instant $t,$
her optimal policy process $\{\tilde{\zeta}_{t}(s),\tilde{c}_{t}(s)\}_{u\in[t,T]}$ is chosen such that
 $$\sup_{\zeta,c}\mathbb{E}\left[\int_{t}^{T}h(u-t)U(c(u)X^{\zeta,c}(u))\,du+h(T-t)\hat{U}(X^{\zeta,c}(T))\right]=$$
$$=\mathbb{E}\left[\int_{t}^{T}h(u-t)U(\tilde{c}(u)X^{\tilde{\zeta},\tilde{c}}(u))\,du+h(T-t)\hat{U}(X^{\tilde{\zeta},\tilde{c}}(T))\right].$$

The value function associated with this stochastic control problem is

\begin{equation}\notag
V(t,s,x)\triangleq\sup_{\zeta,c}\mathbb{E}\left[\int_{s}^{T}h(u-t)U(c(u)X^{\zeta,c}(u))\,du
+h(T-s)\hat{U}(X^{\zeta,c}(T))\bigg|X(s)=x\right],
\end{equation}

$t\leq s\leq T,$ and it solves the following Hamilton-Jacobi-Bellman equation
$$
\frac{\partial V}{\partial s}(t,s,x)+\sup_{\zeta,c}\left[(r+\mu\zeta-c)x\frac{\partial V}{\partial x}(t,s,x)+\frac{1}{2}\sigma^{2}\zeta^{2}x^{2}\frac{\partial^{2} V}{\partial x^{2}}(t,s,x)\right]$$$$+
\frac{h'(s-t)}{h(s-t)}V(t,s,x)+U(xc)=0,
$$
with the boundary condition
\begin{equation}\label{boundarycondition}
V(t,T,x)=\hat{U}(x).
\end{equation}

The first order necessary conditions yield the $t-$optimal policy $\{\tilde{\zeta}_{t}(s),\tilde{c}_{t}(s)\}_{s\in[t,T]}$
\begin{equation}\label{11}
\tilde{\zeta}_{t}(s,x)=-\frac{ \mu\frac{\partial V}{\partial x}(t,s,x)  }{\sigma^{2}x\frac{\partial^{2} V}{\partial x^{2}}(t,s,x)},\quad t\leq s\leq T,
\end{equation}
\begin{equation}\label{c11}
\tilde{c}_{t}(s,x)=\frac{I(\frac{\partial V}{\partial x}(t,s,x))}{x},\quad t\leq s\leq T.
\end{equation}

Therefore, unless the discounting is exponential (in which case $\frac{h'}{h}=\mbox{constant},$ so there is no $t$
dependence in the HJB), the $t-$optimal policy may not be optimal after $t.$ That is
$$ \{\tilde{\zeta}_{t}(s),\tilde{c}_{t}(s)\}_{s\in[t',T]}\neq\argmax_{\zeta, c}\mathbb{E}\left[\int_{t'}^{T}h(u-t')U(c(u)X^{\zeta,c}(u))\,du+h(T-t')\hat{U}(X^{\zeta,c}(T))\right],$$
for some subsequent instant $t',$ so the decision-maker would implement the $t-$optimal policy  at later times 
 only if she is constrained to do so. This failure to remain optimal across times can be regarded as time inconsistency.

Because of time-inconsistency, optimal solutions are irrelevant in practice (although they do exist mathematically) and one must look for an alternative way to solve the problem. This will be done by considering equilibrium policies, that is, policies such that, given that they will be implemented in the future, it is individually optimal to implement them right now. Following \cite{EkeLaz} consider:

\begin{equation}\label{eq}
\bar{\zeta}(s,x)=\frac{F_{1}(s,x)}{x},\quad\bar{c}(s,x)=\frac{F_{2}(s,x)}{x},
\end{equation}
for some functions $F_{1},$ $F_{2}$ and the equilibrium wealth process $\{\bar{X}(s)\}_{s\in[0,T]}$ evolves according to
\begin{equation}\label{dyn}
d\bar{X}(s)=[r\bar{X}(s)+\mu F_{1}(s,\bar{X}(s))-F_{2}(s,\bar{X}(s))]ds+\sigma F_{1}(s,\bar{X}(s))dW(s).
\end{equation} The functions $F_{1}, F_{2}$ are chosen such that on $[t,t+\epsilon]$ it is optimal (this is made precise
in our formal definition of equilibrium policies) to pick
$\bar{\zeta}(t,x)=\frac{F_{1}(t,x)}{x},\quad\bar{c}(t,x)=\frac{F_{2}(t,x)}{x},$ given the agent's wealth at time $t$ is $x,$
and for every subsequent instance $s\geq t+\epsilon$ she follows \eqref{eq}.

\section{ Finite Horizon}

\subsection{General Discount Function}
Let $T$ be a finite time horizon exogenously specified.
In general, for a policy process $\{{\zeta}(s),{c}(s)\}_{s\in[0,T]}$ satisfying \eqref{kj}
 and its corresponding wealth process $\{X(s)\}_{s\in[0,T]}$ (see \eqref{equ:wealth-one})
we denote the expected utility functional
\begin{equation}\label{01FUNCT}
J(t,x,\zeta,c)\triangleq\mathbb{E}\left[\int_{t}^{T}h(s-t)U(c(s)X^{t,x}(s))\,ds+h(T-t)\hat{U}(X^{t,x}(T))\right].
\end{equation}
 We shall give a rigorous mathematical
formulation of the equilibrium policies in the formal definition below.

\begin{definition}\label{finiteh}
A map $F=(F_{1},F_{2}):(0,\infty)\times[0,T]\rightarrow\mathbb{R}\times[0,\infty)$ is an equilibrium policy
for the finite horizon investment-consumption problem, if for any $t,x>0$
\begin{equation}\label{opt}
{\lim_{\epsilon\downarrow 0}}\frac{J(t,x,F_{1},F_{2})-J(t,x,\zeta_{\epsilon},c_{\epsilon})}{\epsilon}\geq 0,
\end{equation}
where
$$J(t,x,F_{1},F_{2})\triangleq J(t,x,\bar{\zeta},\bar{c}),$$
\begin{equation}\label{0eq}
\bar{\zeta}(s)=\frac{F_{1}(s,\bar{X}(s))}{\bar{X}(s)},\quad\bar{c}(s)=\frac{F_{2}(s,\bar{X}(s))}{\bar{X}(s)},
\end{equation}
and $\{\bar{\zeta}(s),\bar{c}(s)\}_{s\in[t,T]}$  should satisfy \eqref{kj}.
 The equilibrium wealth process $\{\bar{X}(s)\}_{s\in[t,T]}$ is a solution of the stochastic differential
equation (SDE)
\begin{equation}\label{0dyn}
d{X}(s)=[r{X}(s)+\mu F_{1}(s,{X}(s))-F_{2}(s,{X}(s))]ds+\sigma F_{1}(s,{X}(s))dW(s).
\end{equation}

The process $\{{\zeta}_{\epsilon}(s),{c}_{\epsilon}(s)\}_{s\in[t,T]}$ is another investment-consumption policy defined by
\begin{equation}\label{1e}
\zeta_{\epsilon}(s)=\begin{cases} \bar{\zeta}(s),\quad s\in[t,T]\backslash E_{\epsilon,t}\\
\zeta(s), \quad s\in E_{\epsilon,t}, \end{cases}
\end{equation}

\begin{equation}\label{2e}
c_{\epsilon}(s)=\begin{cases} \bar{c}(s),\quad s\in[t,T]\backslash E_{\epsilon,t}\\
c(s), \quad s\in E_{\epsilon,t}, \end{cases}
\end{equation}
with $E_{\epsilon,t}=[t,t+\epsilon],$ and  $\{{\zeta}(s),{c}(s)\}_{s\in[t,T]}$ is any
policy for which $\{{\zeta}_{\epsilon}(s),{c}_{\epsilon}(s)\}_{s\in[t,T]}$  satisfy \eqref{kj}.
\end{definition}

The basic idea is the following: the agent (self) has a different rate of impatience $\frac{h'(t)}{h(t)}$ as times goes by
(unless the discounting is exponential) and can be regarded as a continuum of agents (selves);
at every instant $t$ she is building a coalition with her immediate selves $s,$ with $s\in[t,t+\epsilon]$
and try to maximize expected utility of intertemporal consumption and terminal wealth given that the selves on
$[t+\epsilon,T]$ agreed upon an equilibrium strategy.

Our next item in the agenda is to characterize the equilibrium policies by a means of some adjoint processes defined
by a flow (one for every instant $t$) of backward stochastic differential equations (BSDE).
More precisely for every $0\leq t\leq T$ the processes $\{M(t,s), N(t,s)\}_{s\in[t,T]}$ are a solution of the BSDE

\begin{equation}\label{BSDE1}
\begin{cases}
dM(t,s)=-\bigg(M(t,s)\left(r+\mu\frac{\partial{F_{1}}}{\partial{x}}(s,\bar{X}(s))-\frac{\partial{F_{2}}}{\partial{x}}(s,\bar{X}(s))\right)+\sigma \frac{\partial{F_{1}}}{\partial{x}}(s,\bar{X}(s))N(t,s)+\\+h(s-t)\frac{\partial{F_{2}}}{\partial{x}}(s,\bar{X}(s))U'(F_{2}(s,\bar{X}(s)))\bigg)ds+N(t,s)dW(s)\\
M(t,T)=h(T-t)\hat{U}'(\bar{X}(T)),
\end{cases}
\end{equation}
where the equilibrium wealth process $\{\bar{X}(s)\}_{s\in[0,T]}$ follows \eqref{0dyn}.

The next central result does not depend on the choice of the discount function $h.$

\begin{theorem}\label{Hamil1}
Assume there exists a map $F=(F_{1},F_{2}):(0,\infty)\times[0,T]\rightarrow\mathbb{R}\times(0,\infty),$
continuously differentiable with respect to $x$ such that
for every $t\in[0,T]$ there exists a solution  $\{M(t,s), N(t,s)\}_{s\in[t,T]}$
of \eqref{BSDE1} which satisfies
\begin{equation}\label{0au1}
\mu M(t,t)+\sigma N(t,t)=0,
\end{equation}
and
\begin{equation}\label{09p}
 F_{2}(t,x)=I(M(t,t)|X(t)=x).
\end{equation}

Then $F$ is an equilibrium strategy.
\end{theorem}

Appendix A proves this Theorem.

\subsection{Special Discount Functions}

 Following \cite{EkeLaz} we would like to give a characterization of the equilibrium policies
 in terms of both a partial differential equation and an integral equation. We restrict
ourselves to three types of discounting although our method goes far beyond. We consider
exponential, and pseudo-exponential discounting. Let us introduce the Legendre transform of $-U(-x)$
\begin{equation}\label{Leg}
\tilde{U}(y)\triangleq\sup_{x>0}[U(x)-xy]=U(I(y))-yI(y),\quad 0<y<\infty.
\end{equation}
The function $\tilde{U}(\cdot)$ is strictly decreasing, strictly convex and satisfies  the dual relationships
\begin{equation}\label{In12}
\tilde{U}'(y)=-x\quad\mbox{iff}\quad U'(x)=-y.
 \end{equation}

The next Theorem is our main result. It describes the equilibrium policies through a coupled system of parabolic equations.

\begin{theorem}\label{PDE}
Assume there exist two functions $v(t,x)$ and $w(t,x)$ three times continuously differentiable
which satisfy
\begin{equation}\notag
\frac{\partial v}{\partial t}(t,x)+rx\frac{\partial v}{\partial x}(t,x)-
\frac{\mu^{2}}{2\sigma^{2}}\frac{{[\frac{\partial v}{\partial x}}(t,x)]^{2}}{\frac{\partial^{2} v}{\partial x^{2}}(t,x)}+\tilde{U}\left(\frac{\partial v}{\partial x}(t,x)\right) =\alpha_{1j}v(t,x)+\beta_{1j}w(t,x),
\end{equation}
$$
\frac{\partial w}{\partial t}(t,x)+\left(rx-I\left(\frac{\partial v}{\partial x}(t,x)\right)\right)\frac{\partial w}{\partial x}(t,x)
-\frac{\mu^{2}}{\sigma^{2}}\frac{{\frac{\partial v}{\partial x}}(t,x)
{\frac{\partial w}{\partial x}}(t,x)}{\frac{\partial^{2} v}{\partial x^{2}}(t,x)}$$$$+
\frac{\mu^{2}}{2\sigma^{2}}\frac{{[\frac{\partial v}{\partial x}}(t,x)]^{2}\frac{\partial^{2}w}{\partial x^{2}}(t,x)}{[\frac{\partial^{2} v}{\partial x^{2}}(t,x)]^{2}}
=\alpha_{2j}v(t,x)+\beta_{2j}w(t,x),
$$
for all $(t,x)\in[0,T]\times(0,\infty),$
with boundary condition
$$v(T,x)=\hat{U}(x),\qquad w(T,x)=0.$$
Then $F=(F_{1},F_{2})$ given by
 \begin{equation}\label{09con}
 F_{1}(t,x)=-\frac{\mu\frac{\partial v}{\partial x}(t,x)}{\sigma^{2}\frac{\partial^{2} v}{\partial x^{2}}(t,x)},\,\,
F_{2}(t,x)=I\left(\frac{\partial v}{\partial x}(t,x)\right),\,\,\,t\in[0,T],
\end{equation}
is an equilibrium policy. The coefficients $\alpha_{ij},\, \beta_{ij}$ corresponds to different
choices of discount functions. Thus for exponential discounting
$$\alpha_{10}=\delta,\quad \alpha_{20}=0,\quad \beta_{10}=0,\quad \beta_{20}=0,$$
for type I
$$\alpha_{11}=\lambda\rho_{1}+(1-\lambda)\rho_{2},\quad \alpha_{21}=\rho_{1}-\rho_{2},\quad \beta_{11}= 
\lambda(1-\lambda)(\rho_{1}-\rho_{2}),\quad \beta_{21}=\lambda\rho_{2}+(1-\lambda)\rho_{1},$$
and for type II
$$\alpha_{12}=\rho-\lambda,\quad \alpha_{22}=-\lambda,\quad \beta_{12}=\lambda,\quad \beta_{22}=\rho+\lambda.$$
\end{theorem}

Appendix B proves this Theorem.

\begin{remark}
Let us point out that for the case of exponential discounting the equilibrium policy
coincide with the optimal one given by dynamic programming.
Note that the value function $v$
\begin{equation}\label{tr}
 v(t,x)=\sup_{\zeta,c}\mathbb{E}\left[\int_{t}^{T}e^{-\delta(s-t)}U(c(s)X^{\zeta,c}(s))\,ds+
e^{-\delta(T-t)}\hat{U}(X^{\zeta,c}(T))\bigg|X(t)=x\right],
\end{equation}
 and $w=0$ satisfies the PDE system of Theorem \ref{PDE}.
\end{remark}

The next Proposition is in the spirit of Theorem $2$ in \cite{EkeLaz} and gives a description of the equilibrium
policy through an integral equation. It is stated only for the special discounts.

\begin{proposition}\label{IE}
Assume there exist two functions $v(t,x)$ and $w(t,x)$ three times continuously differentiable
which solve the PDE system of Theorem \ref{PDE}. Then $v(t,x)$ satisfy the integral equation
\begin{equation}\label{0ie1}
v(t,x)=\mathbb{E}\left[\int_{t}^{T}h(s-t)U(F_{2}(s,\bar{X}^{t,x}(s)))\,ds+h(T-t)\hat{U}(\bar{X}^{t,x}(T))\right],
\end{equation}
 Recall that $\{\bar{X}(s)\}_{s\in[0,T]}$
is the equilibrium wealth process and it satisfies
\begin{equation}\label{10dyn}
d\bar{X}(s)=[r\bar{X}(s)+\mu F_{1}(s,\bar{X}(s))-F_{2}(s,\bar{X}(s))]ds+\sigma F_{1}(s,\bar{X}(s))dW(s).
\end{equation}
\end{proposition}
Appendix C proves this Proposition.

\begin{remark}
This Proposition suggests that for a general discount function $h(t)$ the equilibrium policies are of
the form \eqref{09con} for a function $v$ as in \eqref{0ie1}. It is not hard to see it holds
for $h(t)=(1+\lambda_{11} t+\lambda_{12}t^{2}+\cdots)\exp(-\rho_{1} t)+(1+\lambda_{21} t+\lambda_{22}t^{2}+\cdots)\exp(-\rho_{2} t).$
\end{remark}

 In the next Proposition stochastic representations for $v$ and $w$ are given.

\begin{proposition}
Assume there exist two functions $v(t,x)$ and $w(t,x)$ three times continuously differentiable
which solve the PDE system of Theorem \ref{PDE}. Then
\begin{equation}\notag
w(t,x)=\mathbb{E}\left[\alpha_{2j}\int_{t}^{T}{\exp(-\beta_{2j}(s-t))}v(s,\bar{X}^{t,x}(s))\,ds\right],
\end{equation}

$$
\!\!\!\!\!\!\!\!\!\!{v}(t,x)=\mathbb{E}\bigg[\int_{t}^{T}{\exp(-\alpha_{1j}(s-t))}\bigg[U\bigg(I\bigg(\frac{\partial v}{\partial x}(s,\bar{X}^{t,x}(s)
\bigg)
\bigg)-$$$$\beta_{1j}w(s,\bar{X}^{t,x}(s))\bigg]\,ds+h(T-t)\hat{U}(\bar{X}^{t,x}(T))\bigg].
$$
Therefore
$${v}(t,x)=\mathbb{E}\bigg[\int_{t}^{T}{\exp(-\alpha_{1j}(s-t))}\bigg[U\bigg(I\bigg(\frac{\partial v}{\partial x}(s,\bar{X}^{t,x}(s)\bigg)\bigg)  -$$ $$
-\alpha_{2j}\beta_{1j}\int_{s}^{T}
\exp(-\alpha_{2j}(z-s))v(z,\bar{X}^{t,x}(z))\,dz\bigg]\,ds+h(T-t)\hat{U}({X}^{t,x}(T))\bigg].$$
\end{proposition}

Proof: It is a direct consequence of Feynman-Kac's formula.
\begin{flushright}
$\square$
\end{flushright}

The main question (since all the results of this section are based on it) is: when does the
system of Theorem \ref{PDE} have solutions? Of course the answer is known for exponential
discounting. Although
there is work in progress, we only have a partial answer at the moment and that is for
CRRA preferences, $U(x)=\hat{U}(x)=\frac{x^{p}}{p}.$ In this case one can disentangle time and
wealth and look for $v(t,x)=f(t)\frac{x^{p}}{p}$ and  $w(t,x)=g(t)\frac{x^{p}}{p},$ where $f(t),$ $g(t)$ solve the ODE system
\begin{equation}\label{on3}
f'(s)+Kf(s)+(1-p)[f(s)]^{\frac{p}{p-1}}= \alpha_{1j}f(s)+  \beta_{1j} g(s),
\end{equation}

\begin{equation}\label{on4}
g'(s)+Kg(s)-pg(s)[f(s)]^{\frac{1}{p-1}}= \beta_{2j}f(s)+   \beta_{2j}g(s),
\end{equation}
for all $t\in[0,T]$ with boundary condition
$$f(T)=1,\qquad  g(T)=0,$$

where $K=rp+\frac{\mu^{2}p}{2\sigma^{2}(p-1)}.$ Existence of this is obvious on small intervals (up to explosions) $[T-\epsilon, T]$ for some $\epsilon>0.$
One can get a global result for small $\beta_{ij}$ (small $\lambda$) by the Implicit Function Theorem.

\section{Infinite Horizon}

We next investigate stationary equilibrium policies and this is done in
the infinite horizon framework. Before engaging into the formal definition
let us point the following key fact.  For a time homogenous
policy process $\{{\zeta}(t),{c}(t)\}_{t\in[0,\infty)}$ satisfying \eqref{kj}
and its corresponding wealth process $\{X(t)\}_{t\in[0,\infty)}$ (see \eqref{equ:wealth-one})
the expected utility functional $J(x,\zeta,c)$ satisfies
\begin{eqnarray}\label{FUNCT}
J(x,\zeta,c)&\triangleq&\mathbb{E}\left[\int_{t}^{\infty}h(s-t)U(c(s)X^{t,x}(s))\,ds\right]
\\\notag
&=&\mathbb{E}\left[\int_{0}^{\infty}h(s)U(c(s)X^{t,x}(t+s))\,ds\right].\\\notag
&=&\mathbb{E}\left[\int_{0}^{\infty}h(s)U(c(s)X^{0,x}(s))\,ds\right].
\end{eqnarray}
This is due to the fact that the processes $\{{X}^{t,x}(t+s)\}_{s\in[0,\infty)}$
and $\{{X}^{0,x}(s)\}_{s\in[0,\infty)}$ have the same $\mathbb{P}$ distribution.
Following the case of finite horizon we have:

\begin{definition}\label{infiniteh}
A map $F=(F_{1},F_{2}):(0,\infty)\rightarrow\mathbb{R}\times[0,\infty)$ is an equilibrium policy
for the infinite horizon investment-consumption problem, if for any $x>0$
\begin{equation}\label{1opt}
{\lim_{\epsilon\downarrow 0}}\frac{J(x,F_{1},F_{2})-J(x,\zeta_{\epsilon},c_{\epsilon})}{\epsilon}\geq 0,
\end{equation}
where
$$J(x,F_{1},F_{2})\triangleq J(x,\bar{\zeta},\bar{c}),$$
\begin{equation}\label{0eq}
\bar{\zeta}(t)=\frac{F_{1}(\bar{X}(t))}{\bar{X}(t)},\quad\bar{c}(t)=\frac{F_{2}(\bar{X}(t))}{\bar{X}(t)},
\end{equation}
and $\{\bar{\zeta}(t),\bar{c}(t)\}_{t\in[0,\infty)}$  should satisfy \eqref{kj}.
 The equilibrium wealth process $\{\bar{X}(t)\}_{t\in[0,\infty)}$ satisfies
\begin{equation}\label{00dyn}
d\bar{X}(t)=[r\bar{X}(t)+\mu F_{1}(\bar{X}(t))-F_{2}(\bar{X}(t))]dt+\sigma F_{1}(\bar{X}(t))dW(t).
\end{equation}

The process $\{{\zeta}_{\epsilon}(t),{c}_{\epsilon}(t)\}_{t\in[0,\infty)}$ is another
time homogenous  investment-consumption policy defined by
\begin{equation}\label{71e}
\zeta_{\epsilon}(t)=\begin{cases} \bar{\zeta}(t),\quad t\in[0,\infty)\backslash E_{\epsilon}\\
\zeta(t), \quad t\in E_{\epsilon}, \end{cases}
\end{equation}

\begin{equation}\label{72e}
c_{\epsilon}(t)=\begin{cases} \bar{c}(t),\quad t\in[0,\infty)\backslash E_{\epsilon}\\
c(t), \quad t\in E_{\epsilon}. \end{cases}
\end{equation}
Here $E_{\epsilon}\subset[0,\infty)$ is a measurable set with
Lebesque measure $|E_{\epsilon}|=\epsilon,$
and  $\{{\zeta}(t),{c}(t)\}_{t\in[0,\infty)}$ is any time homogenous
policy for which $\{{\zeta}_{\epsilon}(t),{c}_{\epsilon}(t)\}_{t\in[0,\infty)}$  satisfies \eqref{kj}.
\end{definition}

In a first step we would like to describe the equilibrium policies in terms of the process $\{M(t), N(t)\}_{t\in[0,\infty)},$
the solution of the infinite horizon BSDE
\begin{equation}\label{BSDE}
\begin{cases}
dM(t)=-(M(t)(\mu F'_{1}(\bar{X}(t))-F'_{2}(\bar{X}(t))+\sigma F'_{1}(\bar{X}(t))N(t)\\
+h(t)F'_{2}(\bar{X}(t))U'(F_{2}(\bar{X}(t)))dt+N(t)dW(t)\\
M(\infty)=0,\quad t\in[0,\infty),
\end{cases}
\end{equation}
and the equilibrium wealth process $\{\bar{X}(t)\}_{t\in[0,\infty)}$ it is given by \eqref{00dyn}.
For more about infinite horizon BSDE see \cite{Chen}. The next result is the infinite
horizon counterpart of Theorem \ref{Hamil1}.

\begin{theorem}\label{Hamil}
Assume there exists a map $F=(F_{1},F_{2}):(0,\infty)\rightarrow\mathbb{R}\times(0,\infty),$
continuously differentiable with respect to $x$ such that there exists a solution  $\{M(t), N(t)\}_{t\in[0,\infty)}$
of \eqref{BSDE} which satisfy
\begin{equation}\label{00au1}
\mu M(0)+\sigma N(0)=0,
\end{equation}
and
\begin{equation}\label{009p}
 F_{2}(x)=I(M(0)|X(t)=x).
\end{equation}

Then $F$ is an equilibrium strategy.
\end{theorem}

The proof is similar to Theorem \ref{Hamil}, so it is skipped.

\begin{flushright}
$\square$
\end{flushright}

As in the preceding Section we find the equilibrium policies for
exponential, type I and type II discounting. With the coefficients
$\alpha_{ij}$ and $\beta_{ij}$  as in Theorem \ref{PDE}, we have
the following formal result.
\begin{theorem}\label{ODE}
Assume there exist two functions $v(x)$ and $w(x)$ three times continuously differentiable
which solve the following ODE system
\begin{equation}\notag
rxv'(x)-\frac{\mu^{2}}{2\sigma^{2}}\frac{v'^{2}(x)}{v''(x)}+\tilde{U}\left(v'(x)\right) =\alpha_{1j}v(x)+\beta_{1j}w(x),
\end{equation}
$$
(rx-I(v'(x)))w'(x)
-\frac{\mu^{2}}{\sigma^{2}}\frac{v'(x)w'(x)}{w''(x)}+
\frac{\mu^{2}}{2\sigma^{2}}\frac{[v'(x)]^{2}w''(x)}{[v''(x)]^{2}}
=\alpha_{2j}v(x)+\beta_{2j}w(x),
$$
for all $x\in(0,\infty).$ If in addition the following transversality
condition
\begin{equation}\label{transv}
M(\infty)=0,
\end{equation}
is satisfied, then $F=(F_{1},F_{2})$ given by
 \begin{equation}\label{009con}
 F_{1}(x)=-\frac{\mu v'(x)}{\sigma^{2}v''(x)},\,\,\,\,
F_{2}(x)=I(v'(x)),
\end{equation}
is an equilibrium policy.
\end{theorem}

The proof follows as in Theorem \ref{PDE}.

\begin{flushright}
$\square$
\end{flushright}

The next Proposition gives the description of equilibrium policies through an integral equation (IE)
as in the Proposition \ref{IE}.

\begin{proposition}\label{1IE}
Assume there exist two functions $v(x)$ and $w(x)$ three times continuously differentiable
which solve the ODE system of Theorem \ref{ODE}. Then $v(x)$ satisfies the integral equation
\begin{equation}\label{9000ie1}
v(x)=\mathbb{E}\left[\int_{0}^{\infty}h(t)U(F_{2}(\bar{X}^{0,x}(t)))\,dt\right].
\end{equation}
\end{proposition}

The proof follows as in Proposition \ref{ODE}.

\begin{flushright}
$\square$
\end{flushright}

\subsection{CRRA Preferences}

In this subsection we are still in the paradigm of exponential,
type I, type II discounts and further investigate the case of $U(x)=\frac{x^{p}}{p}.$
Lets look for the function $v$ of the form $ v(x)=\frac{k{x^{p}}}{p},$ for a constant $k$ which is to be found.
The equilibrium policies are linear in wealth
\begin{equation}\label{lk}
F_{1}(x)=\frac{\mu x}{(1-p)\sigma^{2}},\,\,\,F_{2}(x)=k^{\frac{1}{p-1}}x,
\end{equation}
and the corresponding wealth process is
\begin{equation}\label{as}
\bar{X}(t)=X(0)\exp\left(\left(r+\frac{(1-2p)\mu^{2}}{2(1-p)^{2}\sigma^{2}}-k^{\frac{1}{p-1}}\right)t+ \frac{\mu}{(1-p)\sigma} W(t)\right).
\end{equation}
According to Proposition \ref{1IE} the value function
$v$ should satisfy the integral equation \eqref{9000ie1}
(and this is also sufficient to grant that $F_{1}, F_{2}$ of \eqref{lk} is an equilibrium policy), which in this context becomes
\begin{equation}\label{26}
k^{\frac{1}{1-p}}=\int_{0}^{\infty}h(u)e^{\tilde{k}u}\,du,
\end{equation}
where
\begin{equation}\label{k52}
\tilde{k}=p\left(r+\frac{\mu^{2}}{2(1-p)\sigma^{2}}-k^{\frac{1}{p-1}}\right).
\end{equation}

Let us treat the three cases independently.

\subsubsection{Exponential Discounting}

With $h(t)=\exp(-\delta t),$ \eqref{26} reads

\begin{equation}\label{36}
k^{\frac{1}{1-p}}(\delta-\tilde{k})=1,
\end{equation}

given that
\begin{equation}\label{io1}
\delta>\tilde{k}.
\end{equation}

Thus
\begin{equation}\label{0008}
k=\left[\frac{1}{1-p}\left(\delta-rp-\frac{p\mu^{2}}{2(1-p)\sigma^{2}}\right)\right]^{p-1}.
\end{equation}

The condition

\begin{equation}\label{trans}
\delta>(p\vee0)\left[\frac{\mu^{2}}{2(1-p)\sigma^{2}}+r\right],
\end{equation}

should hold true in view of positivity of $k.$

The adjoint process $\{M(t)\}_{t\in[0,\infty)}$ is given by
$$ M(t)=\exp(-\delta t)v'(\bar{X}(t))$$
$$=kX(0)\left[\exp\left(\left(r(p-1)+\frac{(2p-1)\mu^{2}}{2(1-p)\sigma^{2}}-(p-1)k^{\frac{1}{p-1}}-\delta\right)t+ \frac{\mu}{\sigma} W(t)\right)\right]=$$
$$kX(0)\left[\exp\left(-\left(r+\frac{\mu^{2}}{2\sigma^{2}}\right)t+ \frac{\mu}{\sigma} W(t)\right)\right],$$
whence the condition $M(\infty)=0$ is automatically satisfied.

\begin{remark}
Condition \eqref{trans} is  weaker than the transversality condition
\begin{equation}\label{trans0}
\delta>(p\vee0)\left[\frac{(2-p)\mu^{2}}{2(1-p)\sigma^{2}}+r\right],
\end{equation}
of \cite{Mer69}. If \eqref{trans0} is met the equilibrium policy
coincide with the optimal policy given by dynamic programming.
However if
$$(p\vee0)\left[\frac{\mu^{2}}{2(1-p)\sigma^{2}}+r\right]<\delta\leq (p\vee0)\left[\frac{(2-p)\mu^{2}}{2(1-p)\sigma^{2}}+r\right],$$
equilibrium policy still exists but one cannot prove verification for the optimal policy.
\end{remark}

\subsubsection{ Type I Discounting}

If $h(t)=\lambda\exp(-\rho_{1}t)+(1-\lambda)\exp(-\rho_{2}t),$ equation \eqref{26} leads to

\begin{equation}\label{86}
k^{\frac{1}{1-p}}=\frac{\lambda}{\rho_{1}-\tilde{k}}+\frac{1-\lambda}{\rho_{2}-\tilde{k}},
\end{equation}

given that
\begin{equation}\label{086}
\rho_{i}-\tilde{k}>0,\,\,\, i=1,2.
\end{equation}

 This can be expressed as the positive root of the quadratic equation for $z=k^{\frac{1}{p-1}}$
\begin{equation}\label{7rt}
Q(z)=Az^{2}+Bz+C=0,
\end{equation}
where

\begin{equation}\label{34rt}
A\triangleq 1-p,
\end{equation}

\begin{equation}\label{pk0}
B\triangleq(2p-1)\left(\frac{\mu^{2}}{2(1-p)\sigma^{2}}+r\right)+\frac{\lambda\rho_{2}+(1-\lambda)\rho_{1}}{p}-(\rho_{1}+\rho_{2}),
\end{equation}

\begin{equation}\label{pk1}
C\triangleq-\frac{1}{p}\left(\rho_{1}-rp-\frac{p\mu^{2}}{2(1-p)\sigma^{2}}\right)\left(\rho_{2}-rp-\frac{p\mu^{2}}{2(1-p)\sigma^{2}}\right).
\end{equation}

When $\rho_{1}=\rho_{2}=\delta,$ the equation \eqref{7rt} becomes
\begin{equation}\label{rt0}
\left[(1-p)z-\left(\delta-pr-\frac{p\mu^{2}}{
2(1-p)\sigma^{2}}\right)\right]\left[z+\frac{1}{p}\left(\delta-pr-\frac{p\mu^{2}}{
2(1-p)\sigma^{2}}\right)\right]=0,
\end{equation}
 which compered to \eqref{0008}
brings in a new solution
\begin{equation}\label{artif}
z_{\delta}=\left[-\frac{1}{p}\left(\delta-rp-\frac{p\mu^{2}}{2(1-p)\sigma^{2}}\right)\right].
\end{equation}
 Thus when $\rho_{1},\,\rho_{2}$ are close to $\delta$ the equation \eqref{7rt}
has at least a positive solution provided that $\delta$ satisfies \eqref{trans}. It has two
positive solutions if $p<0.$

Let us search for the transversality condition sufficient to grant  \eqref{transv} ($ M(\infty)=0$).
The process $\{M(t)\}_{t\in[0,\infty)}$ is given by
\begin{equation}\label{1q1}
M(t)\triangleq \lambda\exp(-\rho_{1}t)v'_{1}(\bar{X}(t))+(1-\lambda)\exp(-\rho_{2}t)v'_{2}(\bar{X}(t)),
\end{equation}
where
\begin{equation}\label{00ie1}
v_{i}(x)=\mathbb{E}\left[\int_{t}^{\infty}e^{-\rho_{i}(s-t)}U(F_{2}(\bar{X}^{t,x}_{s}))\,ds\right],\,\,\,i=1,2.
\end{equation}

Consequently
\begin{equation}\label{08}
v_{i}(x)=\frac{k^{\frac{p}{1-p}}x^{p}}{p(\rho_{i}-\tilde{k})},\,\,\,i=1,2.
\end{equation}

Furthermore, for $i=1,2$

$$
\exp(-\rho_{i}s)v'_{i}(\bar{X}(t))$$$$=\frac{X(0)k^{\frac{p}{1-p}}}{(\rho_{i}-\tilde{k})}
\left[\exp\left(\left(r(p-1)+\frac{(2p-1)\mu^{2}}{2(1-p)\sigma^{2}}-(p-1)k^{\frac{1}{p-1}}-\rho_{i}\right)t+ \frac{\mu}{\sigma} W(t)\right)\right].$$

Therefore the transversality condition reads

\begin{equation}\label{o8}
k^{\frac{1}{p-1}}< r+ \frac{1}{1-p}\left[\rho_{i}-\frac{(2p-1)\mu^{2}}{2(1-p)\sigma^{2}}\right],\,\,\, i=1,2.
\end{equation}

In general it is not obvious if the equation \eqref{7rt} has a positive solutions for which the transversality condition
and \eqref{086} are met. However
this is the case when the $\rho_{i}$ are close to $\delta$ and $\delta$ satisfies \eqref{trans}.

  It is interesting to note that in some cases  there may be two equilibrium policies.
  Let us illustrate this in an example: we take $p>\frac{1}{2},\,\, p\simeq\frac{1}{2} $ and denote $y\triangleq r+\frac{\mu^{2}}{
2(1-p)\sigma^{2}}.$ Moreover,
for some small $\epsilon>0$ let $\rho_{1}=p{y}+\epsilon,$  $\rho_{2}=p{y}-\epsilon,$ so that
$B=\frac{\epsilon(1-2\lambda)}{p},$ and $C=\frac{\epsilon^{2}}{p}.$ If $\lambda>\frac{1+\sqrt{4p(1-p)}}{2},  \,\,\lambda\simeq\frac{1+\sqrt{4p(1-p)}}{2}$ the roots of
\eqref{7rt}, $z_{1,2}\simeq\frac{\epsilon}{\sqrt{p(1-p)}}>\frac{\epsilon}{p}$ so that \eqref{086} holds true. The transversality condition
 \eqref{o8} is satisfied if $\epsilon$ is chosen small enough.

\subsubsection{ Type II Discounting}
If $h(t)=(1+\lambda t)\exp(-\rho t),$ equation \eqref{26} leads to

\begin{equation}\label{186}
k^{\frac{1}{1-p}}=\frac{1}{\rho-\tilde{k}}+\frac{\lambda}{(\rho-\tilde{k})^{2}},
\end{equation}

given that
\begin{equation}\label{1086}
\rho-\tilde{k}>0.
\end{equation}

 This can be expressed as the positive root of the quadratic equation for $z=k^{\frac{1}{p-1}}$
\begin{equation}\label{1rt}
Q(z)=Az^{2}+Bz+C=0,
\end{equation}
with

\begin{equation}\label{1pk0}
A\triangleq 1-p,
\end{equation}

\begin{equation}\label{91pk0}
B\triangleq(2p-1)\left(\frac{\mu^{2}}{2(1-p)\sigma^{2}}+r\right)+\frac{\rho(1-2p)}{p}+\frac{\lambda}{p},
\end{equation}

\begin{equation}\label{1pk1}
C=-\frac{1}{p}\left(\rho-rp-\frac{p\mu^{2}}{2(1-p)\sigma^{2}}\right)^{2}.
\end{equation}

The transversality condition in this case is

\begin{equation}\label{1o8}
k^{\frac{1}{p-1}}< r+\frac{1}{1-p}\left[\rho-\frac{(2p-1)\mu^{2}}{2(1-p)\sigma^{2}}\right].
\end{equation}

Similar to the type I discounting it is possible to establish a positive solution
of \eqref{1rt} which satisfy \eqref{1086} and \eqref{1o8}
 when $\lambda$ is close to zero.

\section{Concluding Remarks}

This paper introduced a novel concept in stochastic optimization namely the notion of 
equilibrium policies. We analyze the Merton portfolio management problem in the context
of exponential and non-exponential discounting. Although type I, II and exponential discounting are considered only,
our methodology can be extended for more general discount functions. The optimal policies
are characterized by both an integral equation and a system of partial differential equations.
The infinite horizon case is covered as well and it is shown that in some situations 
there are more equilibrium policies. Moreover for CRRA preferences the equilibrium policies are
to consume and invest in the risky asset a constant proportion of the corresponding wealth,
 which is similar to the optimal policy for exponential discounting, but the constants are different.

\section{Appendix}

{\bf A Proof of Theorem \ref{Hamil1}}

We assume $\hat{U}=r=0$ to ease the presentation.
Let $\{X^{\epsilon}(s)\}_{s\in[0,T]}$ be the wealth corresponding to
$\{{\zeta}_{\epsilon}(s),{c}_{\epsilon}(s)\}_{s\in[0,T]}$ i.e.,
\begin{equation}\label{eq2}
dX^{\epsilon}(s)= X^{\epsilon}(s)((r+\mu\zeta_{\epsilon}(s)-c_{\epsilon}(s))\, ds+\sigma\zeta_{\epsilon}(s)\,dW(s)).
\end{equation}
The processes $\{Y^{\epsilon}(s)\}_{s\in[0,T]}$ and $\{Z^{\epsilon}(s)\}_{s\in[0,T]}$ defined by the SDE
\begin{equation}\label{09}
\begin{cases}
dY^{\epsilon}(s)=Y^{\epsilon}(s)(\mu \frac{\partial F_{1}}{\partial x}(s,\bar{X}(s))-
\frac{\partial F_{2}}{\partial x}(s,\bar{X}(s)))\,ds+
\sigma[Y^{\epsilon}(t)\frac{\partial F_{1}}{\partial x}(s,\bar{X}(s)+\\+(\bar{X}(s)\zeta(s)
-F_{1}(s,\bar{X}(s)){\chi_{E_{\epsilon}}(s)}]\,dW(s)\\
Y^{\epsilon}(0)=0
\end{cases}
\end{equation}
and

$$\begin{cases}
$$dZ^{\epsilon}(s)=[Z^{\epsilon}(s)(\mu \frac{\partial F_{1}}{\partial x}(s,\bar{X}(s))-\frac{\partial F_{2}}{\partial x}(s,\bar{X}(s)))+\\+(\mu\bar{X}(s)(\zeta(s)
-\frac{\partial F_{1}}{\partial x}(s,\bar{X}(s)) )-(c(s)\bar{X}(s)-F_{2}(s,\bar{X}(s)){\chi_{E_{\epsilon}}(s)}]\,ds+$$
\\$$+
[\sigma Z^{\epsilon}(s)\frac{\partial F_{2}}{\partial x}(s,\bar{X}(s)) +(\sigma Y^{\epsilon}(s)(\zeta(s)
-\frac{\partial F_{1}}{\partial x}(s,\bar{X}(s)){\chi_{E_{\epsilon}}(s)}]\,dW(s)
$$\\
$$Z^{\epsilon}(0)=0$$
\end{cases}$$

can be regarded as first order and second order variation of  $\{X(s)\}_{s\in[0,T]}$

At this point we need the following Lemma from \cite{Pen}.

\begin{lemma}\label{Peng}
 For any $k\geq 1$

\begin{equation}\label{1q}
\sup_{s\in[0,\infty]}\mathbb{E}|X^{\epsilon}(s)-\bar{X}(s)|^{2k}=O(\epsilon^{k}),
\end{equation}

\begin{equation}\label{2q}
\sup_{s\in[0,\infty]}\mathbb{E}|Y^{\epsilon}(s)|^{2k}=O(\epsilon^{k}),
\end{equation}

\begin{equation}\label{3q}
\sup_{s\in[0,\infty]}\mathbb{E}|Z^{\epsilon}(s)|^{2k}=O(\epsilon^{2k}),
\end{equation}

\begin{equation}\label{4q}
\sup_{s\in[0,\infty]}\mathbb{E}|X^{\epsilon}(s)-\bar{X}(s)-Y^{\epsilon}(s)|^{2k}=O(\epsilon^{2k})
\end{equation}

\begin{equation}\label{5q}
\sup_{s\in[0,\infty]}\mathbb{E}|X^{\epsilon}(s)-\bar{X}(s)-Y^{\epsilon}(s)-Z^{\epsilon}(s)|^{2k}=o(\epsilon^{2k})
\end{equation}

\end{lemma}

In the light of this Lemma the following expansion holds:

$$
J(t,x,\zeta_{\epsilon},c_{\epsilon})= J(t,x,F_{1},F_{2})$$$$
+\mathbb{E}\int_{t}^{T}\bigg(h(s-t)(U(c(s)\bar{X}^{t,x}(s))-
U(F_{2}(\bar{X}^{t,x}(s))){\chi_{E_{\epsilon}}(s)}$$$$+h(s-t)(Y^{\epsilon}(s)+
Z^{\epsilon}(s))\frac{\partial F_{2}}{\partial x}(s,\bar{X}^{t,x}(s))U'\left(\frac{\partial F_{2}}{\partial x}(s,\bar{X}^{t,x}(s))\right)\bigg)\,ds+o(\epsilon)
$$

In the above equation we would like to get rid of  $\{Y^{\epsilon}(s)\}_{s\in[t,T]},$
$\{Z^{\epsilon}(s)\}_{s\in[t,T]}$ and the way to accomplish this is by using the adjoint
 processes  $\{M(t,s), N(t,s)\}_{s\in[t,T]}$ and integration by parts. It turns out
$$
\!\!\!\!\!\!\!\!\!\!Y^{\epsilon}(s)\frac{\partial F_{2}}{\partial x}(s,\bar{X}(s))U'\left(\frac{\partial F_{2}}{\partial x}(s,\bar{X}(s))\right)\,ds=-Y^{\epsilon}(s)dM(t,s)$$$$-\bigg[Y^{\epsilon}(s)M(t,s)\bigg(\mu\frac{\partial F_{1}}{\partial x}(s,\bar{X}(s))-
\frac{\partial F_{2}}{\partial x}(s,\bar{X}(s)\bigg)
+\sigma Y^{\epsilon}(s) N(t,s)\frac{\partial F_{2}}{\partial x}(s,\bar{X}(s))\bigg]\,ds$$$$+Y^{\epsilon}(s)N(t,s)\,dW(s),
$$

and
$$
\!\!\!\!\!\!\!\!\!\!Z^{\epsilon}(s)\frac{\partial F_{2}}{\partial x}(s,\bar{X}(s))U'\left(\frac{\partial F_{2}}{\partial x}(s,\bar{X}(s))\right)\,ds=-Z^{\epsilon}(s)dM(t,s)$$$$-\bigg[Z^{\epsilon}(s)M(t,s)\bigg(\mu\frac{\partial F_{1}}{\partial x}(s,\bar{X}(s))
-\frac{\partial F_{2}}{\partial x}(s,\bar{X}(s)\bigg)
+\sigma Z^{\epsilon}(s) N(t,s)\frac{\partial F_{2}}{\partial x}(s,\bar{X}(s))\bigg]\,ds$$$$+Z^{\epsilon}(s)N(t,s)\,dW(s),
$$

Consequently

$$\mathbb{E}\int_{t}^{T}h(s-t)(Y^{\epsilon}(s)+Z^{\epsilon}(s))\frac{\partial F_{2}}{\partial x}(s,\bar{X}(s))U'\left(\frac{\partial F_{2}}{\partial x}(s,\bar{X}(s))\right)\,ds=$$$$-\mathbb{E}\int_{t}^{T}(Y^{\epsilon}(s)+Z^{\epsilon}(s))dM(t,s)$$$$\!\!\!\!\!\!\!\!\!\!\!\!\!\!\!\!\!\!-
\mathbb{E}\int_{t}^{T}\bigg[(Y^{\epsilon}(s)+Z^{\epsilon}(s))M(t,s)\bigg(\mu\frac{\partial F_{1}}{\partial x}(s,\bar{X}(s))-\frac{\partial F_{2}}{\partial x}(s,\bar{X}(s)\bigg) +$$$$\sigma (Y^{\epsilon}(s)+Z^{\epsilon}(s))\frac{\partial F_{2}}{\partial x}(s,\bar{X}(s))N(s)\bigg]\,ds.$$

Moreover by \eqref{09} and the equation following it

$$
\mathbb{E}\int_{t}^{T}(Y^{\epsilon}(s)+Z^{\epsilon}(s))M(t,s)\bigg(\mu\frac{\partial F_{1}}{\partial x}(s,\bar{X}(s))-\frac{\partial F_{2}}{\partial x}(s,\bar{X}(s)\bigg)\,ds=$$$$\mathbb{E}\int_{t}^{T}
M(t,s)d(Y^{\epsilon}(s)+Z^{\epsilon}(s))$$$$-\mathbb{E}\int_{t}^{T}[M(t,s)(\mu\bar{X}(s)(\zeta(s) -F_{1}(s,\bar{X}(s)))-(\bar{X}(s)c(s)-F_{2}(s,\bar{X}(s)))]{\chi_{E_{\epsilon}}(s)})\,ds$$

 Thanks to It\^o's formula
\begin{eqnarray}\notag
d(M(t,s)(Y^{\epsilon}(s)+Z^{\epsilon}(s))&=&M(t,s)d(Y^{\epsilon}(s)+Z^{\epsilon}(s))+(Y^{\epsilon}(s)+Z^{\epsilon}(s))dM(t,s)
\\\notag&+&d(Y^{\epsilon}(s)+Z^{\epsilon}(s))dM(t,s),
\end{eqnarray}

which in conjunction with the previous equations yield

$$\!\!\!\!\!\!\!\!\!\!\mathbb{E}\int_{t}^{T}h(s-t)\frac{\partial F_{2}}{\partial x}(s,\bar{X}(s))U'(F_{2}(s,\bar{X}(s))) (Y^{\epsilon}(s)+Z^{\epsilon}(s))\,ds$$$$=\mathbb{E}\int_{t}^{T} [M(t,s)\mu(\bar{X}(s)\zeta(s)-F_{1}(\bar{X}(s))]\,ds$$$$+\mathbb{E}\int_{t}^{T}[(\bar{X}(s)c(s)-F_{2}(\bar{X}(s)))-(
\sigma N(t,s)(\bar{X}(s)\zeta(s)
+F_{1}(\bar{X}(s)))]{\chi_{E_{\epsilon}}(s)}\,ds$$$$-\mathbb{E}\int_{0}^{\infty}\sigma M(t,s)Y^{\epsilon}(s)((\zeta(s)
+F_{1}(\bar{X}(s))){\chi_{E_{\epsilon}}(s)}\,ds.
$$

Next we introduce the Hamiltonian function $H$ by
\begin{equation}\label{Hamiltonian}
H(t,s,x,u,m,n)=mx(\zeta\mu-c)+nx\zeta\sigma+h(s-t)U(cx),\quad u=(\zeta,c).
\end{equation}

The asymptotical expansion leads to

$$
J(t,x,F_{1},F_{2})-J(t,x,\zeta_{\epsilon},c_{\epsilon})=\mathbb{E}\int_{t}^{T}\sigma M(t,s)Y^{\epsilon}(s)\bigg(\frac{\partial F_{2}}{\partial x}(s,\bar{X}^{t,x}(s))
-{\zeta}(s)\bigg){\chi_{E_{\epsilon}}(s)}\,ds$$$$\!\!\!\!\!\!\!\!\!\!+
\mathbb{E}\int_{t}^{T}[H(t,s,\bar{X}^{t,x}(s),\bar{\zeta}(s),\bar{c}(s),M(t,s),N(t,s))$$$$-H(t,s,\bar{X}^{t,x}(s),\zeta(s),c(s),M(t,s),N(t,s))]{\chi_{E_{\epsilon}}(s)}\,ds+o(\epsilon).$$

Additionally, with sufficient integrability assumptions and \eqref{2q}
\begin{equation}\label{2w}
\lim_{\epsilon\downarrow0}\frac{\mathbb{E}\int_{t}^{T}\sigma M(t,s)Y^{\epsilon}(s)\bigg(\frac{\partial F_{2}}{\partial x}(s,\bar{X}^{t,x}(s))
-{\zeta}(s)\bigg){\chi_{E_{\epsilon}}(s)}\,ds}{\epsilon}=0,
\end{equation}

whence

$$\lim_{\epsilon\downarrow0}\frac{J(t,x,F_{1},F_{2})-J(t,x,\zeta_{\epsilon},c_{\epsilon})}{\epsilon}$$
$$=\mathbb{E}\bigg[H\bigg(t,t,x,\frac{F_{1}(t,x)}{x},\frac{F_{2}(t,x)}{x},M(t,t),N(t,t)\bigg)-H(t,t,x,\zeta(t),c(t),M(t,t),N(t,t))\bigg].$$

Therefore a sufficient condition for $F=(F_{1},F_{2})$ to be a equilibrium policy is

\begin{equation}\label{op2}
\left(\frac{F_{1}(t,x)}{x},\frac{F_{2}(t,x)}{x}\right)={\argmax_{\zeta,c}}H(t,t,x,\zeta,c,M(t,t),N(t,t)),
\end{equation}

Finally, the linearity of $H$ in $\zeta$ implies

\begin{equation}\label{10au1}
\mu M(t,t)+\sigma N(t,t)=0,
\end{equation}

and the first order conditions for $c$ (which are also sufficient due to concavity of $U$)

\begin{equation}\label{109p}
 F_{2}(t,x)=I(M(t,t)|X(t)=x).
\end{equation}

\begin{flushright}
$\square$
\end{flushright}

{\bf B Proof of Theorem \ref{PDE}}

The result is established for all three cases (exponential, type I, type II) separately.

{\bf Exponential discounting: $h(t)=e^{-\delta t}$}

Let $v(t,x)$ be the solution of the classical HJB
\begin{equation}\label{0HJB3}
\frac{\partial v}{\partial t}(t,x)+rx\frac{\partial v}{\partial x}(t,x)-\frac{\mu^{2}}{2\sigma^{2}}\frac{{\frac{\partial^{2} v}{\partial x}(t,x)}}{\frac{\partial^{2} v}{\partial x^{2}}(t,x)}+\tilde{U}\left(\frac{\partial v}{\partial x}(t,x)\right)=\delta v(t,x),
\end{equation}
for all $(t,x)\in[0,T]\times(0,\infty),$
with boundary condition
$$v(T,x)=\hat{U}(x).$$

Then $v(t,x)$ and $w(t,x)=0$ solve the parabolic system.

We show the equilibrium strategies are

\begin{equation}\label{0con}
F_{1}(t,x)=-\frac{\mu\frac{\partial v}{\partial x}(t,x)}{\sigma^{2}\frac{\partial^{2} v}{\partial x^{2}}(t,x)},\,\,
F_{2}(t,x)=I\left(\frac{\partial v}{\partial x}(t,x)\right),\,\,\,t\in[0,T].
\end{equation}

Indeed lets consider the processes

\begin{equation}\label{9op0}
M(t,s)=e^{-\delta(s-t)}\frac{\partial v}{\partial x}(s,\bar{X}(s)),\,\, N(t,s)=\sigma e^{-\delta(s-t)} F_{1}(s,\bar{X}(s))\frac{\partial^{2} v}{\partial x^{2}}(s,\bar{X}(s)),
\end{equation}

$s\in[t,T].$ Recall that the equilibrium wealth process $\{\bar{X}(s)\}_{s\in[0,T]},$ is defined by
\begin{equation}\label{10dyn}
d\bar{X}(s)=[r\bar{X}(s)+\mu F_{1}(s,\bar{X}(s))-F_{2}(s,\bar{X}(s))]ds+\sigma F_{1}(s,\bar{X}(s))dW(s).
\end{equation}
It is a matter of direct calculations to prove that  $\{M(t,s), N(t,s)\}_{s\in[t,T]}$ solves BSDE \eqref{BSDE1}. Next
we observe
\begin{equation}\label{10au1}
\mu M(t,t)+\sigma N(t,t)=0,
\end{equation}
and
\begin{equation}\label{109p}
 F_{2}(t,x)=I(M(t,t)|X(t)=x),
\end{equation}
so by Theorem \ref{Hamil1},  $F=(F_{1},F_{2})$ is an equilibrium strategy.

\begin{flushright}
$\square$
\end{flushright}

{\bf{ Type I discounting: \bf{$h(t)=\lambda\exp(-\rho_{1}t)+(1-\lambda)\exp(-\rho_{2}t) $}}}

Let $v$ and $w$ be a solution of the PDE system. We show the equilibrium strategies are

\begin{equation}\label{10con}
F_{1}(t,x)=-\frac{\mu\frac{\partial v}{\partial x}(t,x)}{\sigma^{2}\frac{\partial^{2} v}{\partial x^{2}}(t,x)},\,\,
F_{2}(t,x)=I\left(\frac{\partial v}{\partial x}(t,x)\right),\,\,\,t\in[0,T].
\end{equation}

Let us define the process  $\{M(t,s), N(t,s)\}_{s\in[t,T]}$ by

\begin{equation}\notag
M(t,s)\triangleq \lambda\exp(-\rho_{1}(s-t))\frac{\partial v_{1}}{\partial x}(s,\bar{X}(s)) +(1-\lambda)\exp(-\rho_{2}(s-t))\frac{\partial v_{2}}{\partial x}(s,\bar{X}(s)),
\end{equation}
and
$$
\!\!\!\!\!\!\!\!\!\!\!N(t,s)\triangleq \lambda\sigma\exp(-\rho_{1}(s-t))F_{1}(s,\bar{X}(s))\frac{\partial^{2} v_{1}}{\partial x^{2}}(s,\bar{X}(s))  $$$$+(1-\lambda)\sigma\exp(-\rho_{2}(s-t))F_{1}(s,\bar{X}(s))\frac{\partial^{2} v_{2}}{\partial x^{2}}(s,\bar{X}(s)),
$$

for some functions $v_{1}$ and $v_{2}$ which will be specified later on and $\{\bar{X}(s)\}_{s\in[0,T]}$ is the equilibrium wealth process of \eqref{10dyn} .
By requesting  $\{M(t,s), N(t,s)\}_{s\in[t,T]}$ to solve BSDE \eqref{BSDE1}, one finds a PDE system for $(v_{1},v_{2}).$

Indeed, on one hand  by It\^o's formula

$$dM(t,s)=\bigg(\lambda\exp(-\rho_{1}(s-t))\bigg[-\rho_{1}\frac{\partial v_{1}}{\partial x}(s,\bar{X}(s))
+\frac{\partial^{2} v_{1}}{\partial x\partial s}(s,\bar{X}(s))
$$$$+(\mu F_{1}(s,\bar{X}(s))-F_{2}(s,\bar{X}(s)))\frac{\partial^{2} v_{1}}{\partial x^{2}}(s,\bar{X}(s))+
\frac{\sigma^{2}}{2}F^{2}_{1}(s,\bar{X}(s))\frac{\partial^{3} v_{1}}{\partial x^{3}}(s,\bar{X}(s))\bigg]
$$$$+(1-\lambda)\exp(-\rho_{2}(s-t))\bigg[-\rho_{2}\frac{\partial v_{2}}{\partial x}(s,\bar{X}(s))+\frac{\partial^{2} v_{2}}{\partial x\partial s}(s,\bar{X}(s))
+$$$$+(\mu F_{1}(s,\bar{X}(s))-F_{2}(s,\bar{X}(s)))\frac{\partial^{2} v_{2}}{\partial x^{2}}(s,\bar{X}(s))
+\frac{\sigma^{2}}{2}F^{2}_{1}(s,\bar{X}(s))\frac{\partial^{3} v_{2}}{\partial x^{3}}(s,\bar{X}(s)) \bigg]\bigg)ds$$$$+N(t,s)dW(s).$$

On the other hand from \eqref{BSDE1}

$$dM(t,s)=\bigg(\lambda\exp(-\rho_{1}(s-t))\bigg[\frac{\partial v_{1}}{\partial x}(s,\bar{X}(s))\left(\mu\frac{\partial{F_{1}}}{\partial{x}}(s,\bar{X}(s))-\frac{\partial{F_{2}}}{\partial{x}}(s,\bar{X}(s)\right)+$$$$+\sigma^{2}F_{1}(s,\bar{X}(s))\frac{\partial{F_{1}}}{\partial{x}}(s,\bar{X}(s))\frac{\partial^{2} v_{1}}{\partial x^{2}}(s,\bar{X}(s))+\frac{\partial{F_{2}}}{\partial{x}}(s,\bar{X}(s))U'(F_{2}(s,\bar{X}(s)))   \bigg]+$$
$$+(1-\lambda)\exp(-\rho_{2}(s-t))\bigg[\frac{\partial v_{2}}{\partial x}(s,\bar{X}(s))\left(\mu\frac{\partial{F_{1}}}{\partial{x}}(s,\bar{X}(s))-\frac{\partial{F_{2}}}{\partial{x}}(s,\bar{X}(s)\right)+$$$$+\sigma^{2}F_{1}(s,\bar{X}(s))\frac{\partial{F_{1}}}{\partial{x}}(s,\bar{X}(s))\frac{\partial^{2} v_{2}}{\partial x^{2}}(s,\bar{X}(s))+\frac{\partial{F_{2}}}{\partial{x}}(s,\bar{X}(s))U'(F_{2}(s,\bar{X}(s)))\bigg]\bigg)ds+$$  $$+N(t,s)dW(s).$$

The two representations of $dM(t,s)$ yield

$$-\rho_{1}\frac{\partial v_{1}}{\partial x}(s,x)+\frac{\partial^{2} v_{1}}{\partial x\partial s}(s,x) +(\mu F_{1}(s,x)-F_{2}(s,x))\frac{\partial^{2} v_{1}}{\partial x^{2}}(s,x)$$$$+\frac{\sigma^{2}}{2}
F^{2}_{1}(s,x)\frac{\partial^{3} v_{1}}{\partial x^{3}}(s,x) = -\left(\mu \frac{\partial F_{1}}{\partial x}(s,x)-\frac{\partial F_{2}}{\partial x}(s,x)\right)\frac{\partial v}{\partial x}(s,x)
$$$$-\sigma^{2}F_{1}(s,x)\frac{\partial F_{1}}{\partial x}(s,x)\frac{\partial^{2} v_{1}}{\partial x^{2}}(s,x)  -  \frac{\partial F_{2}}{\partial x}(s,x)\frac{\partial v}{\partial x}(s,x),$$

and

$$-\rho_{2}\frac{\partial v_{2}}{\partial x}(s,x)+\frac{\partial^{2} v_{2}}{\partial x\partial s}(s,x) +(\mu F_{1}(s,x)-F_{2}(s,x))\frac{\partial^{2} v_{2}}{\partial x^{2}}(s,x)$$$$+\frac{\sigma^{2}}{2}
F^{2}_{1}(s,x)\frac{\partial^{3} v_{2}}{\partial x^{3}}(s,x)= -\left(\mu \frac{\partial F_{1}}{\partial x}(s,x)-\frac{\partial F_{2}}{\partial x}(s,x)\right)\frac{\partial v_{2}}{\partial x}(s,x)
$$$$-\sigma^{2}F_{1}(s,x)\frac{\partial F_{1}}{\partial x}(s,x)\frac{\partial^{2} v_{2}}{\partial x^{2}}(s,x)  -  \frac{\partial F_{2}}{\partial x}(s,x)\frac{\partial v}{\partial x}(s,x).$$

This can be rewritten as
$$\frac{\partial}{\partial x}\bigg[\frac{\partial v_{1}}{\partial s}(s,x)-\rho_{1}v_{1}(s,x)
+(\mu F_{1}(s,x)-F_{2}(s,x))\frac{\partial v_{1}}{\partial x}(s,x)$$$$+
\frac{\sigma^{2}}{2}
F^{2}_{1}(s,x)\frac{\partial^{2} v_{1}}{\partial x^{2}}(s,x)+U(F_{2}(s,x)) \bigg]=0,$$
and
$$\frac{\partial}{\partial x}\bigg[\frac{\partial v_{2}}{\partial s}(s,x)-\rho_{2}v_{2}(s,x)
+(\mu F_{1}(s,x)-F_{2}(s,x))\frac{\partial v_{2}}{\partial x}(s,x)$$$$+
\frac{\sigma^{2}}{2}
F^{2}_{1}(s,x)\frac{\partial^{2} v_{2}}{\partial x^{2}}(s,x)+U(F_{2}(s,x)) \bigg]=0.$$

Recall that $v$ and $w$ is a solution of the PDE system. Thus
\begin{equation}\label{9ei0}
v_{1}\triangleq v+(1-\lambda)w,\qquad v_{2}\triangleq v-\lambda w
\end{equation}
satisfy the above PDEs. Moreover
\begin{equation}\label{110au1}
\mu M(t,t)+\sigma N(t,t)=0,
\end{equation}
and
\begin{equation}\label{1109p}
 F_{2}(t,x)=I(M(t,t)|X(t)=x),
\end{equation}
so by Theorem \ref{Hamil1},  $F=(F_{1},F_{2})$ is an equilibrium strategy.

\begin{flushright}
$\square$
\end{flushright}

{\bf{ Type II discounting: \bf{$h(t)=(1+\lambda t)\exp(-\rho t) $}}}

Let $v$ and $w$ be a solution of the PDE system. We show the equilibrium strategies are

\begin{equation}\label{10con}
F_{1}(t,x)=-\frac{\mu\frac{\partial v}{\partial x}(t,x)}{\sigma^{2}\frac{\partial^{2} v}{\partial x^{2}}(t,x)},\,\,
F_{2}(t,x)=I\left(\frac{\partial v}{\partial x}(t,x)\right),\,\,\,t\in[0,T].
\end{equation}

Let us define the process  $\{M(t,s), N(t,s)\}_{s\in[t,T]}$ by

\begin{equation}\notag
M(t,s)\triangleq \exp(-\rho(s-t))\frac{\partial v_{1}}{\partial x}(s,\bar{X}(s)) +\lambda t\exp(-\rho(s-t))\frac{\partial v_{2}}{\partial x}(s,\bar{X}(s)),
\end{equation}
and
$$
\!\!\!\!\!\!\!\!\!\!\!N(t,s)\triangleq \sigma\exp(-\rho_{1}(s-t))F_{1}(s,\bar{X}(s))\frac{\partial^{2} v_{1}}{\partial x^{2}}(s,\bar{X}(s))  $$$$+\sigma\lambda t \exp(-\rho_{2}(s-t))F_{1}(s,\bar{X}(s))\frac{\partial^{2} v_{2}}{\partial x^{2}}(s,\bar{X}(s)),
$$

for the functions
\begin{equation}\label{71ei0}
v_{1}\triangleq v,\qquad v_{2}\triangleq v- w.
\end{equation}

 As for the case of type I one can check that $\{M(t,s), N(t,s)\}_{s\in[t,T]}$  solves BSDE \eqref{BSDE1}. Moreover
\begin{equation}\label{1110au1}
\mu M(t,t)+\sigma N(t,t)=0,
\end{equation}
and
\begin{equation}\label{11109p}
 F_{2}(t,x)=I(M(t,t)|X(t)=x),
\end{equation}
so by Theorem \ref{Hamil1},  $F=(F_{1},F_{2})$ is an equilibrium strategy.

\begin{flushright}
$\square$
\end{flushright}

{\bf C Proof of Proposition \ref{IE}}

 If the discounting is exponential the result is a direct consequence of Feynman-Kac's formula. For type I,
 the functions $v_{1}$ and $v_{2}$
 of \eqref{9ei0} admit the following stochastic representations

\begin{equation}\label{000ie1}
v_{i}(t,x)=\mathbb{E}\left[\int_{t}^{T}e^{-\rho_{i}(s-t)}U(F_{2}(\bar{X}^{t,x}_{s}))\,ds\right],\,\,\,i=1,2,
\end{equation}
by Feynman-Kac's formula. Consequently
\begin{equation}\label{0000ie1}
v(t,x)=\lambda v_{1}(t,x)+(1-\lambda )v_{2}(t,x)
=\mathbb{E}\left[\int_{t}^{T}h(s-t)U(F_{2}(\bar{X}^{t,x}_{s}))\,ds\right].
\end{equation}
Similarly for type II, \eqref{0000ie1} holds true.

\begin{flushright}
$\square$
\end{flushright}

{\bf{Acknowledgements}}

\vspace{0.4cm}

The authors would like to thank Ali Lazrak for helpful discussions and comments.

\end{document}